# Self-Powered Broadband Photodetector Based on $MoS_2/Sb_2Te_3$ Heterojunctions: A promising approach for highly sensitive detection

Hao Wang[1,2], Yaliang Gui[1], Chaobo Dong[1], Salem Altaleb[1], Behrouz Movahhed Nouri[2], Martin Thomaschewski[1], Hamed Dalir[1,2], and Volker J. Sorger[1,2]

[1]Department of Electrical and Computer Engineering, George Washington University, Washington, DC 20052, USA

[2]Optelligence LLC, Upper Marlboro, Maryland 20772, USA

**Abstract**

Topological insulators have shown great potential for future optoelectronic technology due to their extraordinary optical and electrical properties. Photodetectors, as one of the most widely used optoelectronic devices, are crucial for sensing, imaging, communication, and optical computing systems to convert optical signals to electrical signals. Here we experimentally show a novel combination of topological insulators (TIs) and transition metal chalcogenides (TMDs) based self-powered photodetectors with ultra-low dark current and high sensitivity. The photodetector formed by a $MoS_2/Sb_2Te_3$ heterogeneous junction exhibits a low dark current of 2.4 pA at zero bias and 1.2 nA at 1V. It shows a high photoresponsivity of > 150 mA $W^{-1}$ at zero bias and rectification of 3 times at an externally applied bias voltage of 1V. The excellent performance of the proposed photodetector with its innovative material combination of TMDs and TIs paves the way for the development of novel high-performance optoelectronic devices. The TIs/TMDs transfer used to form the heterojunction is simple to incorporate into on-chip waveguide systems, enabling future applications on highly integrated photonic circuits.

**Introduction**

Transition metal chalcogenides (TMDs) have been widely explored as active materials for high-performance optoelectronics due to their unique physical properties.[1–3] Transition metal chalcogenides optoelectronics usually suffer from surface oxidation, high contact resistance, and relatively low mobilities.[4–7] To alleviate the aforementioned disadvantages of TMD-based photodetectors, recent technologies utilize plasmonic enhancement to improve the light-matter interaction of TMDs, while others use heterostructures to adjust inherent electrical and optical characteristics and graphene contact to aid charge carrier injection and extraction; those all exhibit attractive performance improvements such as higher sensitivity, broader frequency response, improved external quantum efficiencies (EQE), higher detectivity (D*) and a broader spectral photoresponse.[4–9] However, combining topological insulators (TIs) as a novel quantum materials group with the attractive features of TMDs has remained largely unexplored. Topological insulators itself exhibit various appealing optoelectronic features[12–14] which find applications in high-performance photodetection, e.g., due to its unique energy band structure. Furthermore, the topologically protected gapless conductive edge states or surface states in TIs offer extraordinarily high mobility and a broad detection spectrum compared to graphene which has zero bandgaps.[15–19] Topological insulators are also mechanically flexible materials that can be used in wearable technology. Most TMDs are easily oxidized under atmospheric environments, whereas TIs are more stable in normal conditions. As a monolayer, molybdenum disulfide ($MoS_2$) has a direct bandgap of 1.8 eV, whereas, in bulk, it has an indirect bandgap of 1.3 eV.[20] The visible to the infrared absorption spectrum of $MoS_2$ ranges from 350 to 950 nm which is utilized in various optoelectronic applications such as light-harvesting, photovoltaics, and photodetection.[21–26] Here, we demonstrate a vertical heterojunction-based photodetector with TIs and TMDs, which provide a promising tool for manipulating the electrical and optical properties of the individual materials. The heterojunction-based photodetector, formed by $MoS_2$ and $Sb_2Te_3$ exfoliation and deterministic transfer to a silicon oxide substrate, shows reduced dark current, high responsivity R=$I_{photo}/P_{laser}$ (defined by the ratio between the photocurrent $I_{photo}$ and the incident laser power $P_{laser}$), and a self-driven detection range from visible to near-infrared wavelengths. The photocarrier separation is accelerated, and photocarrier recombination is suppressed due to the inherent difference in their energy level. Compared to single TMDs, it also exhibits a reduced dark current, which is beneficial for achieving high detection sensitivities at reduced power consumption. The dark current is ~10 pA, and the light-induced on/off ratio is more than $10^3$ due to the interlayer built-in field and a broadband response from 500 nm to 900 nm. Besides the demonstrated free-space application, the proposed device has potential in densely integrated applications such as digital to analogy convertor system or photonic integrated circuits due to the ease of integration with photonic waveguide technology. [23–26]

**Results and Discussion**

To fabricate the device, $Sb_2Te_3$ and $MoS_2$ layers were prepared by mechanical exfoliation and transferred to a $SiO_2$/Si substrate by the pick and drop technique. These two layers are stacked by weak van der Waals force (Fig.1a). Three Ti/Au electrodes formed by electron beam evaporation act as the electrical contact pads to inject and extract the photocurrent generated by the photodetector. Two electrodes are placed on the top of the $MoS_2$ and one electrode is placed on the $Sb_2Te_3$ (Fig.1b). The compact total junction size is measured to be only 0.16 mm$^2$. A Schottky barrier is present when the electrode is contacted with the $MoS_2$/$Sb_2Te_3$ layers to reach the equilibrium state, and it also generates a tunneling current. However, the applied heterostructure minimizes the depletion region due to the atomically thin layer. The photocarriers can be efficiently separated at the heterostructure interface with the built-in electric field in the PN heterojunction or an external electric field from an applied bias, yielding photocurrent across the heterostructure channel. As a result, there is no significant potential barrier in the forward direction, which increase the photodetector's responsivity (Fig.1c).

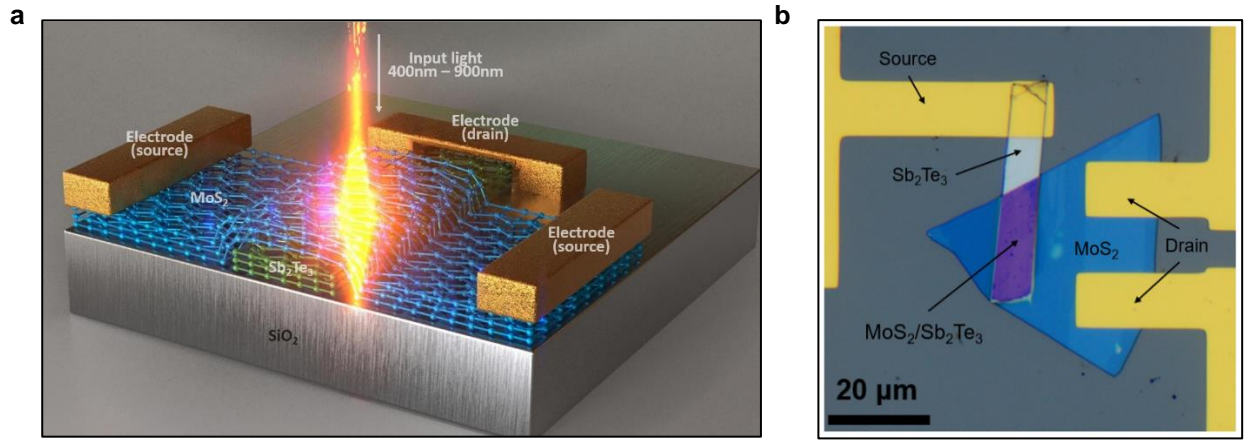

**Figure 1. $Sb_2Te_3$/$MoS_2$ pn junction heterostructure photodetector** (a) A schematic representation of the $Sb_2Te_3$/$MoS_2$ van der Waals p-n junction photodetector. (b) The optical microscope image of PN junction device

Raman and Energy-dispersive X-ray spectroscopy (EDS) measurements are used to characterize the spatial structure and properties of the vertically stacked $MoS_2$/$Sb_2Te_3$ system. Raman spectra are collected via 532 nm laser illumination. The Raman signal of $MoS_2$ showed two distinct peaks at 384 cm$^{-1}$ and 409 cm$^{-1}$ in the spectrum Fig. 2(d), which were indicated the in-plane Mo-S phonon mode ($E^1_{2g}$ at 384 cm$^{-1}$) and out of plane Mo-S phono mode ($A_{1g}$ at 409 cm$^{-1}$) of $MoS_2$ respectively. [30,31] Raman Spectra evidence the $MoS_2$ multilayer film with a stoke shift of 25 cm$^{-1}$.[32] The Raman spectrum of $Sb_2Te_3$ consists of different peaks related to $Sb_2Te_3$ vibrations: 70 cm$^{-1}$, 114 cm$^{-1}$, 130 cm$^{-1}$ with two peaks centered at 97 cm$^{-1}$, and 105 cm$^{-1}$ between, 146 cm$^{-1}$, 167 cm$^{-1}$. The 70 cm$^{-1}$, 97 cm$^{-1}$ and 167 cm$^{-1}$ peaks indicated the $A_{1g}$ and $E_g$ normal modes of the Sb-Te vibrations. The 114 cm$^{-1}$, 105 cm$^{-1}$, and 146 cm$^{-1}$ indicate the Te-Te interactions, which together consist of the whole structure of $Sb_2Te_3$.[32] The thickness of $Sb_2Te_3$ is estimated to be above 65nm.[33] Energy Dispersive Spectroscopy (EDS) is employed to explore the element information collected from the single layers (the exfoliated $MoS_2$ (point C), and $Sb_2Te_3$ (point A) and stacked layers ($MoS_2$/$Sb_2Te_3$ (point B)) on the device, respectively (Fig. 2(b)). The corresponding $MoS_2$ EDS spectrum shows the L series peaks of Mo and S $K_\alpha$ peak is very pronounced; therefore, only a widened peak with a tail towards the high energy end can be seen with EDS. The elemental compositions (Sb/Te) of the exfoliated $Sb_2Te_3$ layers were also determined by EDS. We observed two weak X-ray emission peaks corresponded to Sb and Te. The spectrum also showed the characteristic peaks of Si, which originates from the used substrate. The absence of any other peaks indicates that the $Sb_2Te_3$ layers are formed from Sb and Te, only. The quantitative atomic ratio of Sb and Te is roughly 38.7% to 61.3%, which is near to the 2:3 stoichiometry ratio in accordance to the EDS investigations of $Sb_2Te_3$.[34–36]

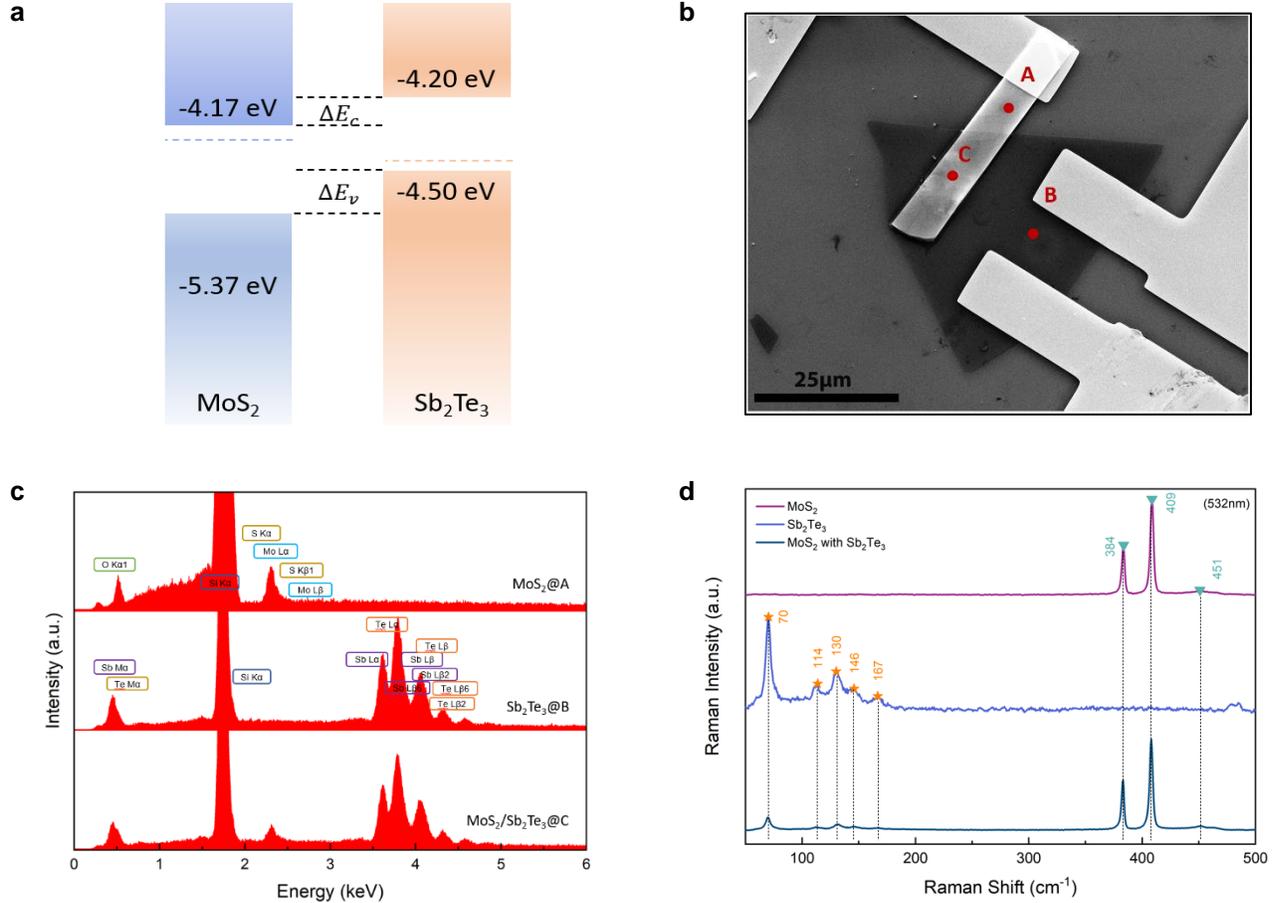

**Fig. 2. Material characterization of the Sb2Te3/MoS2 heterojunction** (a) Band structures of the vdW layered $MoS_2/Sb_2Te_3$ heterojunction (b) SEM image of the fabricated device with point A on $Sb_2Te_3$, point B on $MoS_2$ and point C on the $MoS_2/Sb_2Te_3$ heterojunction. Two electrodes are on top of the $MoS_2$ layer and one on top of the $Sb_2Te_3$. (c) As shown in the SEM image, EDS signals were collected at three distinct positions to characterize the materials. (d) Raman spectra were collected from the pure $MoS_2$ and $Sb_2Te_3$ and the Sb2Te3/MoS2 PN junction region with a 532 nm laser.

Prior to the optoelectronic measurements, we characteristic the electrical performance of the $MoS_2/Sb_2Te_3$ junction by applying a source-drain bias voltage $V_{sd}$ from -1 to 1 V to evaluate the dark current and performance of the van der Waals PN heterojunction. As indicated by the black curve in the I-V measurement (inset of Fig. 3a, 3b and 3c), the dark currents are stable and consistent. The dark current is as low as 2.38 pA at zero bias and 1.16 nA at 1 V bias, which benefits from the heterostructure design. To our best knowledge, this is the lowest dark current in a vdW heterojunction photodetector compared to other vdW heterojunction photodetectors.[37–45] At a positive bias, the dark current becomes significantly higher than at negative voltages. Furthermore, under on and off bias voltage $V_{sd}$, the source-drain current, $I_{sd}$ has a rectifying characteristic. This reveals that a vdW PN heterojunction is created at the interface of the stacked $MoS_2/Sb_2Te_3$. The diode's on/off ratio, defined as the ratio between photogenerated current and dark current, is around $5.6 \times 10^3$. At room temperature, the ultra-low dark current and high on/off ratio at zero bias enable the development of a highly sensitive self-powered photodetector. The power dependent IV measurement of the PN heterojunction at the wavelength of 500 nm, 700 nm, and 900 nm has also been performed (inset of Fig. 3a, 3b and 3c). At zero bias voltage, the heterojunction separates the photogenerated electro-hole pairs and introduces the photocurrent due to the built-in electric field formed at the $MoS_2/Sb_2Te_3$ interface. The photocurrent $I_{photo}$ increases with higher optical power illumination because electro-hole pairs generation rates increase. This demonstrates self-powered photodetection at various wavelengths (Fig. 3a, 3b and 3c with the insets in Figure 3(a)(b)(c) showing the measured $I_{photo}$ and $I_{sc}$ as a function of the optical power at various wavelengths. The $I_{photo}$ increases proportionally to the illuminated optical power, which gives evidence of the photovoltaic effect in the vdW PN heterojunction. The responsivity of the Sb$_2$Te$_3$/MoS$_2$ PN junction-based heterostructure photodetector is measured from -1 V to +1 V at the wavelength of 500 nm, 700 nm, and 900 nm using the free-space optical setup. With a positive bias voltage, an external electric field is formed at the junction interface, increasing the separation efficiency of the photogenerated carriers and, therefore, the responsivity. At +1 V bias, the responsivity was amplified 1.5 to 3 times depending on the wavelength and optical power. The responsivity grows with the applied voltage and gradually saturates depending on the incident optical power. Under zero bias, the significant broadband photocurrent of the PN junction photodetector was observed within a wavelength range from 500 to 900 nm (Fig.3d). The highest responsivity under zero bias was 156 mA W$^{-1}$ at 650 nm.

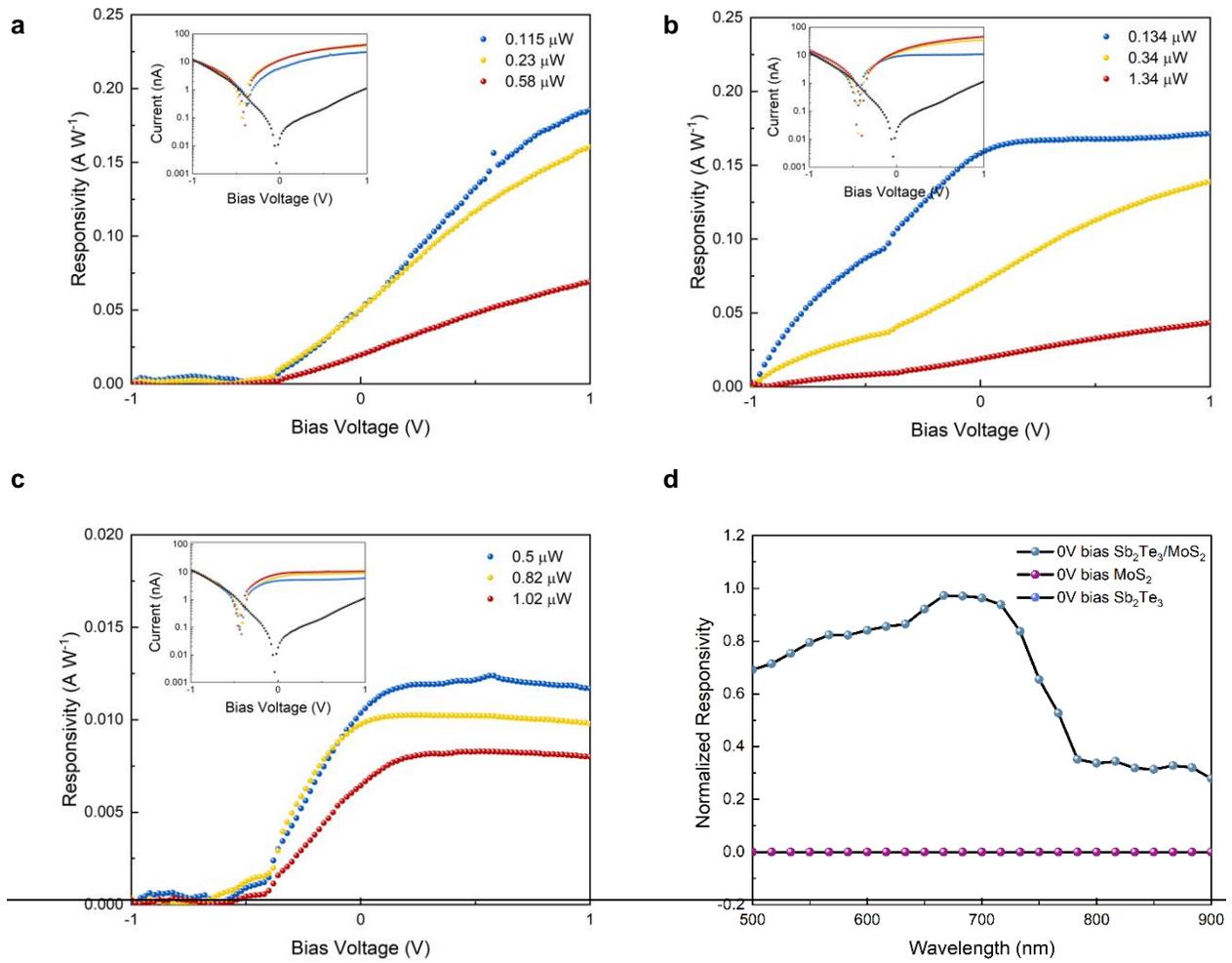

**Fig. 3. Electrical characterization of Sb$_2$Te$_3$/MoS$_2$ PN junction heterostructure photodetector.** (a), (b), (c) The measured photoresponsivity at different wavelengths, 500nm, 700nm, and 900nm, under -1 to 1 V bias voltage. At zero bias, the photoresponsivity was clearly observed and increased proportionally to the optical power. (d) Normalized responsivity by sweeping the wavelength under 0 bias from 500 nm to 900 nm with a step size of 10nm, which indicates a broad wavelength response.

## Conclusion

In conclusion, we proposed and experimentally demonstrate a self-powered, extremely sensitive photodetector based on a novel combination of a topological insulator and TMDs van der Waals PN heterojunction. Due to the current rectifying property of the PN heterojunction, the constructed device exhibits extremely low dark current and high responsivities. Under zero bias, the dark current of the device is 2.4 pA and the responsivity of 156 mA W$^{-1}$. By increasing the bias voltage to 1 V, the responsivity was magnified thrice due to the external field amplification. This outstanding performance of the proposed photodetector with its novel material combination paves the way for further research and applications based on other TMDs and Tis in optoelectronic applications.[46–49] The PN heterojunction transfer can be easily integrated into waveguide systems, allowing for future applications in PICs. Indeed, as nano-optics becomes practical[50], efficient photodetectors augment other optoelectronic devices based on 2D materials[51] such novel sources or energy-converters for renewable energy[52].

## Methods

### Device fabrication

The $Sb_2Te_3/MoS_2$ PN junction heterostructure is formed using 2D flakes exfoliated from the bulk crystals and transferred by the pick and drop transfer system (See supplementary for more information) on $Si/SiO_2$ substrate. The electrical contact pads were formed using e-beam lithography (Raith Pioneer EBL) and metal thermal evaporation were employed to deposit the Ti/Au (5 nm / 50 nm) electrodes on the produced heterostructures. The lift-off was performed by acetone, then rinsing in isopropyl alcohol and nitrogen drying in RT (room temperature).

### Device experimentation

The tunable (NKT SUPERCONTINUUM Compact) source and the source meter (Keithly 2600B) was used for electrical response measurements of $Sb_2Te_3/MoS_2$ PN junction heterostructure devices. The laser beam was focused on the devices by an objective lens. The Raman was performed using a 532 nm laser source at room temperature.

## Acknowledgments

This work was performed in part at the George Washington University Nanofabrication and Imaging Center (GWNIC).

layered 2D materials. *Nanotechnology* **26**, (2015).